\documentclass[prb,aps,preprint,superscriptaddress]{revtex4}
\usepackage{graphicx}
\usepackage{amssymb}
\usepackage{bm}

\begin{document}

\author{Toshifumi Itakura}
 \affiliation{Department of Physics, Nagoya University, Nagoya 464-8602, Japan}
 \email{itakurat@s6.dion.ne.jp}

\title{Paraelectric-to-ferroelectric crossover and electron dynamics
from time-dependent Hartree-Fock calculations}

\begin{abstract}
In recent years, researchers have discovered various excited states of
electrons having
 an excited domain with a structural order different from the original one.
We examine a dynamics of electron system and phonon system
by using the time-dependent Hartree-Fock approximation.
We examine time evolution with spin degree
of freedom. 
We found a thermally excited paraelectric-to-ferroelectric crossover.
The metastable state is spin pierls state.
We also examine the Landau expansion of electron system at quarter-filled.
\end{abstract}

\pacs{78.47.+p 71.30.+h 71.38.-k 78.40.Me}

\maketitle

The paraelectric-to-ferroelectric crossover caused by electrons displacement
is induced by an electron 
in an paraelectric insulating solid when 
it is excited by a electric external field.
\cite{Nasu}
The original concept of this relaxation assumes 
this crossover to be a microscopic phenomenon.
The photo induced paraelectric-to-ferroelectric crossover
can be called a photo-induced structural phase transition.
In this paper, we examine the theoretical studies on photo induced problem.
Photo-induced phase transitions have attracted much attention.
This structural phase transition has often been described as a domino effect.
It is highly nonlinear and cooperatives,
in particular its time evolution is many body problem.
The photo-irradiation a large energy is 
imported by the photo induced irradiation that is not accessible in 
thermal equilibrium.
Thus even if photo-induced phase is one that is realized in such conventional
method, the mechanism of the transition is not trivial at all.
A large enhancement of dielectric constant by UV light irradiation is
reported for the first time in quantum paraelectric family of
perovskite oxides, such as SrTiO$_3$ and KTiO$_3$,
under a weak DC field.
\cite{Koshihara}
The photo irradiation induce
the paraelectric-to-ferroelectric crossover.
as a function of temperature.
Without photo irradiation, the dielectric property of these material is
Quantum paraelectric.
The DC electric field suppress the quantum fluctuation,
which enhance the quantum domino catastrophes.

Photo-induced transitions have been studied in many method.
Whether a deterministic or stochastic approach accurately describes 
the time evolution of electron system would depend 
on the time scale of observation.
The recent study obviously needs to be described in a deterministic manner.
\cite{Yonemitsu}
Their charge and lattice motion is calculated by using
the time-dependent Schrodinger equation.
In this paper, we study another itinerant electron system and phonon system.
The present study differs from other studies,
because we treat the electron transportation and many body interaction.

We examined the time evolution from the initial state to a stationary state 
of the electron system with random initial condition.
During the evolution, the excited state changes into a metastable state.
This method has advantage because in Ref. \cite{Yonemitsu},
the spin degree of freedom can not be treated.
Though, in our method, we can not treat the eigenstate
in stead of that we can examine  time evolution with spin degree
of freedom. 
We also examine the Landau expansion of electron system at quater-filled.

The electron system and classical phonon system
 is described by a time dependent Hartree-Fock  approximation.
 The classical phonon system is evaluated by leap-flog method.
We assume a that Hubbard model for thermal-fluctuation-induced 
paraelectric-to-ferroelectric crossover, as follows, 
\begin{eqnarray}
H_{el} &=& \Sigma_{i,\sigma}  [
h_{i}^F ( c_{i,\sigma}^{\dagger} c_{i+1,\sigma}
  + h.c. ) 
  + h_{i,\sigma}^H n_{i,\sigma} ], \\
 h_{i}^F & = & t_0 + \alpha ( u[i+1] - u[i] ),   
 h_{i,\sigma}^H = \frac{U}{2} n_{\bar{\sigma}}[i,t], \\
 H_{ph} &=& \sum_i \frac{P_i^2}{2 M} + k u[i]^2 
\end{eqnarray}
where $t_0$ is tight binding hopping strength,
 and $U$ is on-site Coulomb repulsion.
$c_{i, \sigma}^{\dagger}, c_{i, \sigma}$ is creation and annihilation operator
at cite $i$, spin $\sigma$.
$H_{ph}$ phonon system Hamiltonian, $P_i$ is momentum, $M$ is mass
$k$ is stiffness, $u[i]$ is distortion 
and $\alpha$ is dimerization coupling constant between electron system
and phonon system.
We will examine the simple Hubbard model.
While, the complexity of degrees of freedom are different,
the universality should be the same,
because the important quantity is the single band electron system.
The time evolution is described by the correlation function,
\begin{equation}
g [ i,j,t ]^{\sigma \sigma'}
=<c_{i,\sigma}^{\dagger} (t)  c_{j,\sigma'} (t) >.
\end{equation}
The time evolution equations of a spin diagonal correlation function are
\begin{eqnarray}
 \label{eqn:electron}
\frac{ d  g [ i, j ,t]^{\sigma}   }{d t} 
   &=& -\frac{i}{\hbar} [
   h_{i}^F g [ i + 1 , j , t]^{\sigma}  
   + h_{i-1}^F g [ i-1, j,t ]^{\sigma}  \nonumber \\
 &-& h_{j-1}^F 
 g [ i , j-1,t ]^{\sigma}   
 - h_{j}^F g [ i , j + 1,t]^{\sigma}    \nonumber \\
 &+& h_{i,\sigma}^H  g [ i ,j,t ]^{\sigma}  
 - h_{j,\sigma}^H  g [i, j,t] ^{\sigma}   ].
 \nonumber \\
 \end{eqnarray}
However, there is no eigenstate for this correlation function,
in stead we can examine the time trace of system with random initial
condition numerically.
The  equation corresponds to the random phase approximation.
In present study, d-dimensional quantum field theory is equivalent to
2d classical field theory.
During the time evolution, the initial excited state
changes into the  local ground state.

The Bloch-Winger representation of the correlation function is,
\begin{equation}
g^{\sigma} [x,p] = \sum_m  e^{i m p}  g^{\sigma} [k,l] ,
\end{equation}
where $x=k-l$, $m=(k+l)/2$.
This representation is not positive definite, 
therefore probabilistic interpretation can not be applied.

First, we examine numerical solution.
The numerical method comprises the Crank-Nicholson method
\cite{NC}
and the Gauss-Seigel Method.
The time evolution of the Schrodinger equation is described by the unitary 
condition to 2nd-order approximation: 
\begin{equation}
{\rm exp} ( - i H t ) \simeq \frac{1 + i \frac{1}{2} H t}{ 1 - i \frac{1}{2} H t}.
\end{equation}
In present numerical study, we set the number of electrons to be 4
due to quarter-filled system.
Let us examine the time dependence of the electron order parameter.
CDW  is the intensity of the alternative  derivative
 of local charge density.
AF  is the intensity of the alternative  derivative
 of local spin density.
The order parameters are defined by
\begin{eqnarray}
{\rm CDW} &=& \sum_{i,\sigma} (-1)^i  (
   g[i+1,i+1,t]^{\sigma}+g[i-1,i-1,t]^{\sigma} \nonumber \\
  &-& 2 
  g[i,i,t]^{\sigma}), \nonumber \\
  {\rm AF} &=& \sum_{i,\sigma} \sigma (-1)^i  (
   g[i+1,i+1,t]^{\sigma}+g[i-1,i-1,t]^{\sigma} \nonumber \\
  &-& 2 
  g[i,i,t]^{\sigma}). \nonumber \\
\end{eqnarray}
For the whole system the physical parameter is as follows,
$t_0 = 0.1, 
U   = 0.5$, $k=8$, $M=500$, $\alpha=0.1$ 
and unit of the time is normalized to the above parameters.  
The initial conditions are 
$ g[i,j,t]^{\sigma} =0.25 + \beta*n$
where $n$ is complex random value with absolute value 1 for all $i,j.\sigma$.
and  are $\beta_{el}=0.01$ or $\beta_{el}=0.1$.
This means the initial condition is a quarter-filled homogeneous state 
with off-diagonal coherence.
For the Bloch-Wigner representation, after the time evolution,
the momentum distribution of the charges 
is oscillating and has domain in real space.
The strengths of higher moment are weak.
This result is similar to the Haldane conjectures.

\begin{figure}[tb]
\begin{center}
\unitlength 1mm
\begin{picture}(90,90)(0,5)
\put(0,0){\resizebox{85mm}{!}{\includegraphics{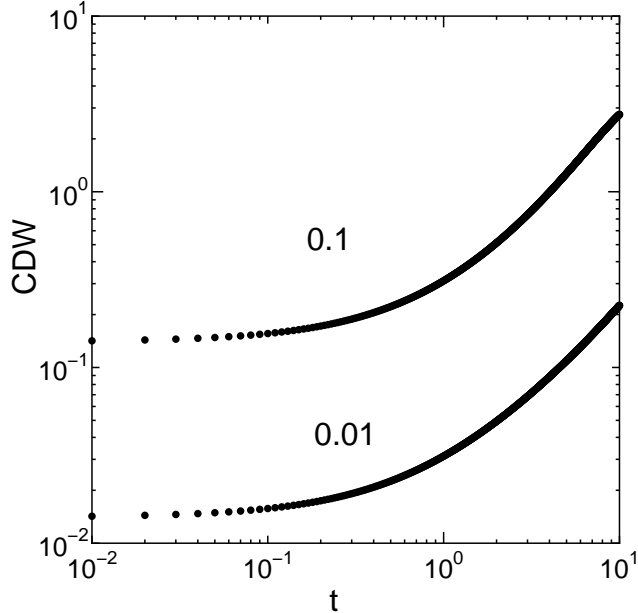}}}
\end{picture}
\end{center}
\caption{\label{fig:exact} Time dependence of electron order parameter.
$\beta_{el}=0.01$and $0.1$ no phonon}
\end{figure}

\begin{figure}[tb]
\begin{center}
\unitlength 1mm
\begin{picture}(90,90)(0,5)
\put(0,0){\resizebox{85mm}{!}{\includegraphics{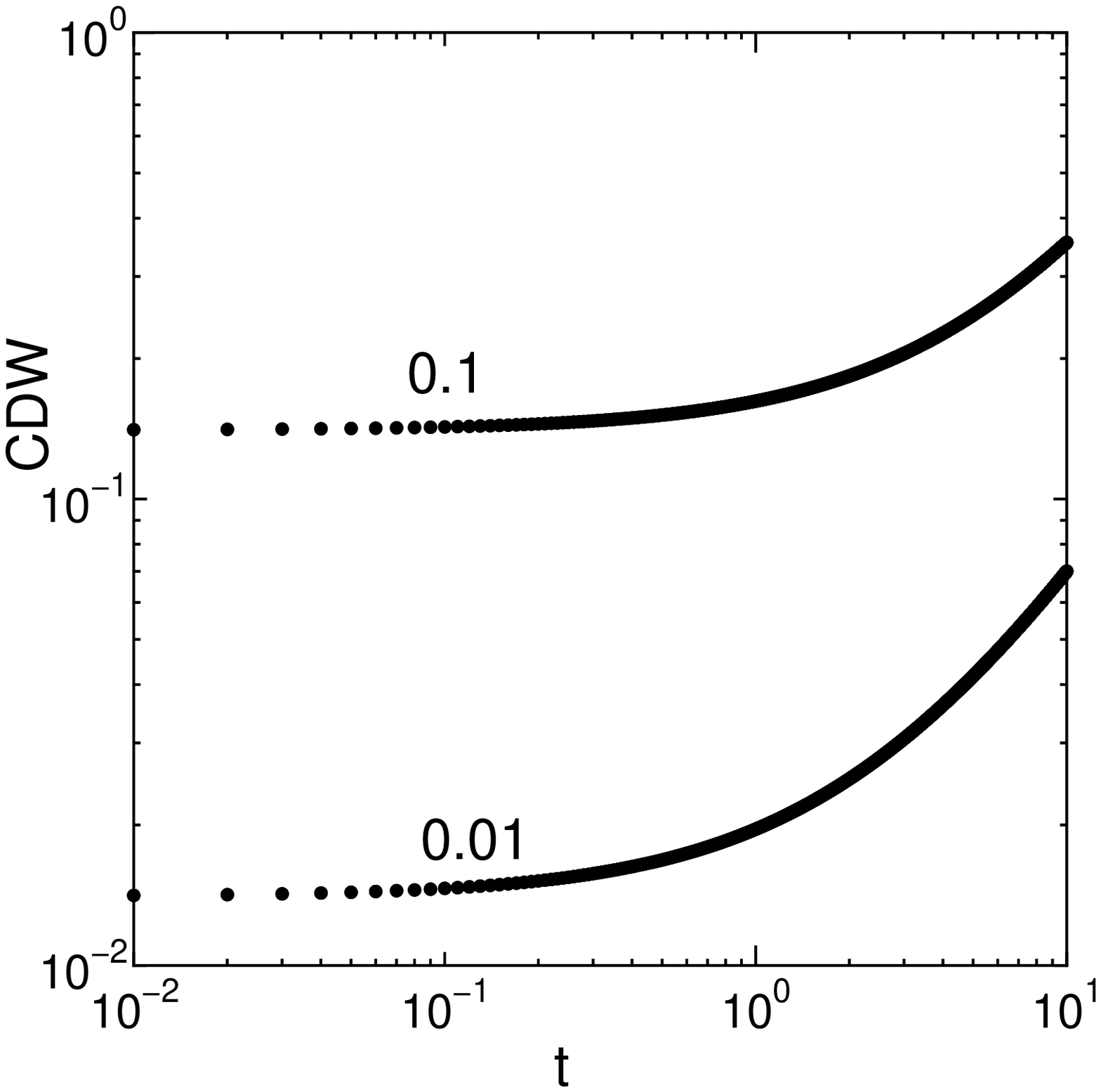}}}
\end{picture}
\end{center}
\caption{\label{fig:exact} Time dependence of electron order parameter.
$\beta_{el}=0.01$ and $0.1$ aco phonon}
\end{figure}

\begin{figure}[tb]
\begin{center}
\unitlength 1mm
\begin{picture}(90,90)(0,5)
\put(0,0){\resizebox{85mm}{!}{\includegraphics{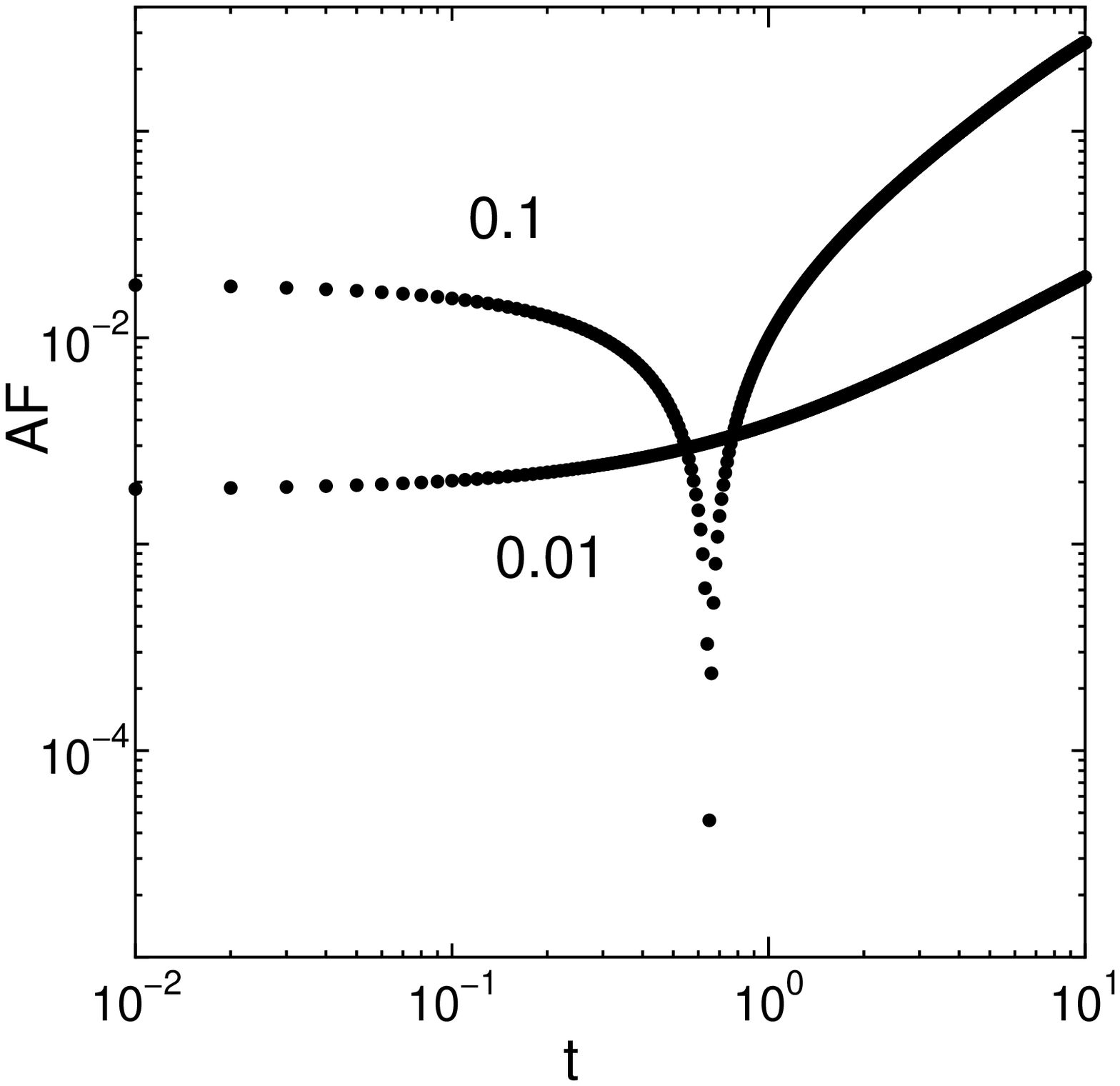}}}
\end{picture}
\end{center}
\caption{\label{fig:exact} Time dependence of spin order parameter.
$\beta_{el}=0.01$and $0.1$ no phonon}
\end{figure}

\begin{figure}[tb]
\begin{center}
\unitlength 1mm
\begin{picture}(90,90)(0,5)
\put(0,0){\resizebox{85mm}{!}{\includegraphics{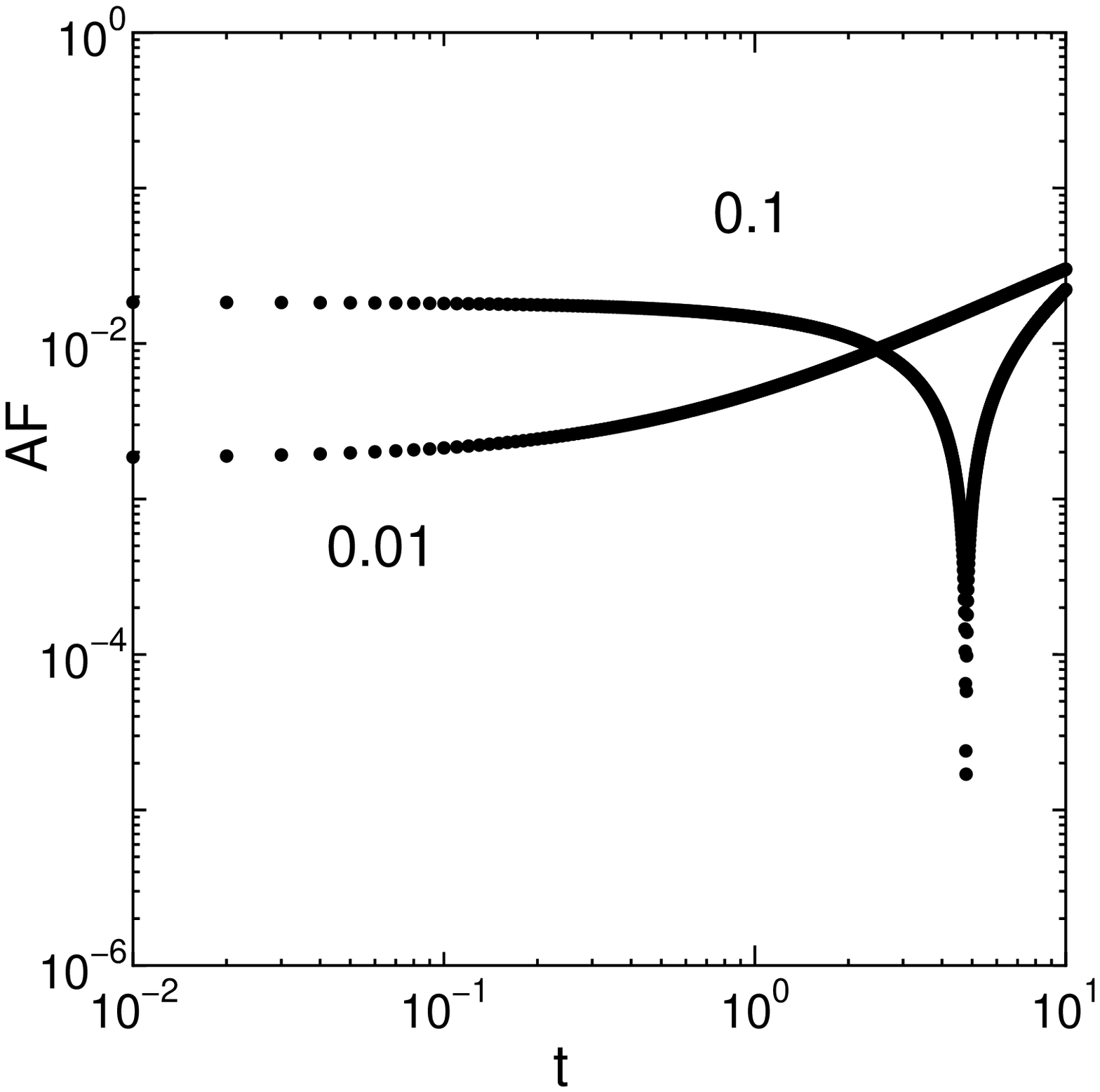}}}
\end{picture}
\end{center}
\caption{\label{fig:exact} Time dependence of spin order parameter.
$\beta_{el}=0.01$ and $0.1$ aco phonon}
\end{figure}

The CDW state order parameter is $4k_F$ CDW order parameter.
\cite{Seo}
The AF state order parameter is $4k_F$ ant ferromagnetic order parameter.
Thus, the 4$k_F$ CDW is quantum paraelectric order parameter,
The dimerization enhance the 4$k_F$ CDW state.
The AF state order parameter represent spin pierls instability.
For strong noise, AF state order parameter decreases and
 CDW state order parameter increases with time.
The time trace of  the two order parameters become in phase.
The time trace of order parameters shows the phase of two order parameters
depends on noise strengths.
Exactly the final stage, CDW state order parameter is positive for the figs.1 
and 2,
and AF state order parameter phase is become negative for figs.3 and 4, 
when strong randamness. 
This phase locking due to 
the quasi-one-dimensional interchain coupling is assumed.
The figs.1 and 2 is ground state and figs. 3 and 4 is metastable state.
The strong randomness leads to metastable state.

Due to complex initial condition, the transition from metastable state to
the order ground state occurs.
When $\beta$=0.01, the AF order parameter increases,
this  indicates the Wigner crystal insulator.
When $\beta$=0.1, the AF order parameter become negative,
the cross over to spin pierls state (ferroelectric) occurs.
An experiment has shown the existence of
a photo-induced paraelectric-to-ferroelectric phase crossover.
\cite{Koshihara}
Thus, we have examine this crossover from the viewpoint of
non-equilibrium condensed matter theory by using a Hubbard model Hamiltonian
in view point of time dependent Hartree-Fock calculation.
The transverse photo effect behaves 
as thermal fluctuation.
The thermal fluctuation creates a  non-equilibrium state
with finite order parameter and
settled random initial condition.
The complex random initial condition correspond to
the excited phonon with imaginary self energy.
Thus the emition and absorption of phonon system by electric external  field
is essential for photo-induced effect.
An external electric field leads to the catastrophe for the domino system.
The paraelectric comes from quantum fluctuations between dominoes.
This just the thermal fluctuation induced quantum domino effect.
The displacement of the dipole is suppressed 
because of the repulsion between the dipole moment.

Next, we divide system A and B sub lattice.
The Bloch-Wigner representation has to generate effective energy.
First, the property of Liuville operator is given by,
\begin{equation}
\frac{d g^{\sigma} [x,p]}{d t} 
= \frac{1}{\hbar} \{W_{\sigma}, g^{\sigma} [x,p] \},
\end{equation}
\begin{equation}
\{ A , B \} =  \frac{\partial A}{\partial x} \frac{\partial B}{\partial p}
              - \frac{\partial A}{\partial p} \frac{\partial B}{\partial x}, 
\end{equation}
\begin{equation}
W_{\sigma} = -2t_0 \cos p + h_x^{H}.
\end{equation}              
The obtained effective energy is derived Moyal-Krammer formulation,
as follows,
\begin{equation}
E_{eff} = \sum_{\sigma,\alpha} W_{\sigma,\alpha} n_{\sigma,\alpha}.
\end{equation}
where $\alpha$ and A or B.
Present representation of effective energy 
\begin{eqnarray}
 E_{eff} &=& 
  \sum_{\sigma} ( \frac{U}{2} <n_{\bar{\sigma}}^A> 
  - 2 t_0 \cos p  ) <n_{\sigma}^A>. \nonumber \\
 &&  \sum_{\sigma} ( \frac{U}{2} <n_{\bar{\sigma}}^B> 
  - 2 t_0 \cos p  ) <n_{\sigma}^B>. 
\end{eqnarray}
where we neglect the fock term.
Next we assume the system is Wigner crystal.
The charge density is 4k$_F$ and spin density is non magnetic.
\begin{equation}
< n_{\sigma}^A > = \frac{1 + \Delta }{4},
< n_{\sigma}^B > =\frac{1 - \Delta}{4}.
\end{equation} 
Thus, effective energy is given by
\begin{equation}
E_{eff}= U \frac{1-\Delta^2}{8} - 2 t_0 \cos p
\end{equation}
Then we examine Landau expansion which obtained by free energy,
\begin{equation}
 F_{eff} = E_{eff}|_{\Delta=0} + \frac{\delta E_{eff}}{\delta \Delta} \Delta
 + \frac{1}{2} \frac{\delta^2 E_{eff}}{\delta \Delta^2}  \Delta^2
 + k_B T \int dx dp [f \ln f + (1-f) \ln (1-f) ].
\end{equation}
The last part of the above equation is approximated as follows,
\begin{eqnarray}
 && k_B T  \int dp dx [ \ln \frac{f}{1-f} |_{\Delta=0} 
 \frac{\delta f}{\delta \Delta}|_{\Delta=0} \Delta \nonumber \\
 &+& k_B T \int dp dx [ \frac{1}{f} + \frac{1}{1-f} ]|_{\Delta=0}
 ( \frac{\delta f}{\delta \Delta})^2 |_{\Delta=0}
  \frac{\Delta^2}{2} \nonumber \\
 &+& k_B T  \int dp dx [ \ln \frac{f}{1-f} ]|_{\Delta=0} 
  \frac{\delta^2 f}{\delta \Delta^2}|_{\Delta=0}
   \frac{\Delta^2}{2} \nonumber \\
 &+& k_B T  
 \int dp dx [ -\frac{1}{f^2} + \frac{1}{(1-f)^2} ]_{\Delta=0}
 (\frac{\delta f}{\delta \Delta} )^3 |_{\Delta=0}
  \frac{\Delta^3}{3!} \nonumber \\
 &+& k_B T  \int dp dx [\frac{1}{f} + \frac{1}{1-f}] |_{\Delta=0}
 \frac{\partial}{\partial \Delta}|_{\Delta=0}
  { (\frac{\partial f}{\partial \Delta})^2}|_{\Delta=0}
 \frac{\Delta^3}{3!} \nonumber \\
 &+& k_B T  
 \int dx dp [\frac{2}{f^3} + \frac{2}{(1-f)^3}] |_{\Delta=0}
 ( \frac{\partial f}{\partial \Delta})^4 |_{\Delta=0}
  \frac{\Delta^4}{4!} \nonumber \\
 &+& k_B T  
 \int dp dx [ -\frac{1}{f^2} + \frac{1}{(1-f)^2} ] |_{\Delta=0}
 \frac{\partial^4 f}{\partial \Delta^4} |_{\Delta=0} \frac{\Delta^4}{4!}
 \nonumber \\
 &+& k_B T  
 \int dp dx [ -\frac{1}{f} + \frac{1}{(1-f)} ] |_{\Delta=0}
 \frac{\partial^2 }{\partial \Delta^2} |_{\Delta=0}
  (\frac{\partial f}{\partial \Delta})^2 |_{\Delta=0}
  \frac{\Delta^4}{4!}. \nonumber \\
\end{eqnarray}
In summary 2-nd order expansion is
\begin{eqnarray}
A_2 &=& \frac{U}{4} (1 - \frac{U}{4} \rho (e_F)),
\end{eqnarray}
where $ \rho (\epsilon_F) $ is density of state of electron.
The 3-rd order coefficient is
\begin{eqnarray}
A_3 &=& \frac{k_B T \rho (e_F) U}{4}.  
\end{eqnarray}
The expression of 4-th order is given as
\begin{eqnarray}
A_4 &=& \frac{k_B T \rho (e_F) U}{4}.
\end{eqnarray}
Thus, this system is asymmetric double well system.
However, due to the low dimensional fluctuation
 there are no phase transition.
 
In summary, we found a thermally excited paraelectric-to-ferroelectric
 crossover.
We examined time evolution of electron system with spin degree of freedom
and phonon system.  
The quantum noise corresponds phonon excited by the electric external  field.
The imaginably random initial condition
 corresponds to emition and absorption of excited phonon
by electric external field.
The stronger randomness leads to metastable spin pierls state.
The metastable state is unstable non-equilibrium  state.
This is ferroelectric state.
We suggest such a non-equilibrium effect is necessary 
to photo induced effects.
We also examine the Landau expansion of electron system at quarter-filled.

\end{document}